\documentstyle[preprint,epsf]{jpsj}
\def \virg{\;\;,}
\def \point{\;\,.}
\def \kf{k_{\rm F}}
\def \vf{v_{\rm F}}
\def \e{{\rm e }}
\def \lb0{\overline{l_0}}
\def\ggs{\buildrel\textstyle > \over {\hbox{\raise0.2ex\hbox{$\sim$}}}}
\def\lls{\buildrel\textstyle < \over {\hbox{\raise0.2ex\hbox{$\sim$}}}}
\def\gsim{\,\lower0.75ex\hbox{$\ggs$}\,}
\def\lsim{\,\lower0.75ex\hbox{$\lls$}\,}
\def\runtitle{
On the Mott glass 
  in  the one-dimensional half-filled charge density waves
  }
\def\runauthor
{
 Yoshikazu {\sc Suzumura} 
 and   
 Masato {\sc Isobe}   
}

\title{
On the Mott glass 
  in  the one-dimensional half-filled charge density waves   
}

\author{
Yoshikazu  {\sc Suzumura} 
\footnote{E-mail: e43428a@nucc.cc.nagoya-u.ac.jp}
   and 
  Masato {\sc Isobe}    
\footnote{E-mail: isobe@slab.phys.phys.nagoya-u.ac.jp}
}

\inst{
Department of Physics, Nagoya University, Nagoya 464-8602 \\ 
}

\recdate{\hspace{35mm}}

\abst
{
We study the  effect of impurity pinning on a
 one-dimensional half-filled electron system, which is expressed 
 in terms of   a phase Hamiltonian with   
  the charge degree of freedom. Within  the classical treatment,
 the pinned state is examined  numerically.
 The  Mott glass,
 which has been pointed out by Orignac {\it et al.}
[Phys. Rev. Lett {\bf 83} (1999)2378], 
 appears  in the intermediate region where   
 the impurity potential competes with   the commensurate potential. 
Such a state is verified 
   by calculating   the   soliton formation energy, 
the local restoring force  around the   pinned state    
and the optical  conductivity.    
}

\kword
{
 Impurity pinning, Mott glass, commensurability pinning, 
   charge fluctuation, soliton formation,  optical conductivity 
  }

\begin{document}
\sloppy
\maketitle
\section{Introduction}
 In one-dimensional electron systems, Anderson 
 localization corresponding to 
 the  impurity pinning 
  takes place  even for  weak impurities.
\cite{Impurity1,Impurity2,localization}
 The pinning  is enhanced  for repulsive interactions, as found in 
 the charge density waves 
\cite{Suzumura_Fukuyama_JPSJ84,Giamarchi_PRB1988}
 where the spatial variation 
 of   the phase spreads widely due to randomness.
\cite{Fukuyama_Lee}  
However, such a pinning becomes complicated for   
 the system with  both the  commensurate band filling and  the  repulsive 
 interaction, 
 in which the phase is also pinned by  
  the commensurate  potential,
\cite{Fukuyama_JPSJ78}
 i.e, umklapp scattering.
It has been maintained that  
 the Mott insulator with a charge excitation gap occurs 
  for large commensurate potential and 
     the compressive Anderson glass with no excitation gap
       appears for the weak random impurities.
\cite{Fujimoto_Kawakami_PRB1996,Mori_Fukuyama_JPSJ1996}
Such a competition has been further 
 investigated by Orignac,  Giamarchi  and 
 Doussal (OGD),
\cite{Orignac_PRL} 
 who applied the variational method to the replicas for 
 the impurity potential. 
 For  the case of the impurity potential 
  being much larger than the commensurate potential, 
   one obtains the Anderson glass, in which     
    the optical conductivity exhibits  $\sigma (\omega) \propto \omega^2$ 
 i.e., with no significant effect of the  commensurate potential.
The opposite case leads to the  Mott insulator where 
 the compressibility becomes zero and the gap appears in 
 the optical conductivity.
However  there exists  the intermediate  case  called 
   the Mott glass where  
  the  compressibility becomes zero 
 but  the conductivity shows the gapless behavior 
 similar to that of the Anderson glass.
\cite{Orignac_PRL,Fujimoto_JPSJ2000}
 Such a state comes from two  kinds of length scales ;  
 one  is the localization  length 
 and the other is a width of the kink followed 
 by the soliton formation. 
The vanishing of the  compressibility  indicates 
  a finite energy to create  the soliton of the phase of 
 the charge density wave, i.e.,  
 the relevance  of the  commensurate potential.
 Thus the energy for the soliton is determined mainly by the local 
 region around the kink. 
On the other hand,  the absence of the gap in the conductivity 
 comes from the property of the impurity pinning  with  
   the localization length, which is  different from that of the kink.
 Thus these two kinds of pinning could be compatible 
 and  lead to a certain region of coexistence.

Such a coexistence is explored in   the present paper 
 by examining the properties of 
 the ground state and the excited state of a  system,   
 which  lead to  the competition between 
 the impurity potential and the commensurate potential.
  We study numerically 
 a  half-filled electron system with repulsive interaction 
 based on the phase Hamiltonian. 
For the simplicity, we focus on the charge degree of freedom.
Based on our preliminary work,
\cite{Suzumura_Meeting}
 we demonstrate  that such a Mott glass state does appear 
   with increasing the commensurate potential as  
  the crossover from  the impurity pinning to the commensurability 
 pinning.   
In \S2, formulation is given 
 and the pinned states are examined to determine 
 the competition between the Anderson glass and the Mott glass.
In \S3, the competition between the Mott glass and the Mott insulator 
  is examined by calculating the local fluctuation 
 around the pinned state, e.g., 
   the restoring force 
and the optical conductivity.
Discussion is given in \S4. 

\section{Model and the Pinned  State}
The formulation for calculating  a model   numerically 
  is given  within the classical treatment.
 The parameter  for  
  the  competition between the impurity potential and commensurate 
potential  is derived. 
 The excitation energy  for the soliton formation is calculated 
in the presence of  impurity where 
 the appearance of the finite energy is identified with 
 a  crossover from the Anderson glass to the Mott glass.

\subsection{Model}
 We consider a one-dimensional Hamiltonian  with a length $L$ given by
\cite{Suzumura_Fukuyama_JPSJ84} 
\begin{eqnarray}
 \label{eq:Hamiltonian_cont}
{H}
 =\int^L_0 dx \left[A\left(\frac{d\theta}{dx}\right)^2
   -V\sum^{N_{imp}}_{j=1}\cos(\theta+2k_Fx)\delta(x-R_j)
     -B\cos(2\theta)\right]  \virg
\end{eqnarray}
 where $\theta(x)$
 denotes a phase variable of the charge density wave.
\cite{Suzumura_PTP79}
 The $V$ term denotes the impurity potential where  
 $N_{imp}$ is the total number of the impurity and 
   $R_j$ is the location of the impurity at  the $j$-th site. 
Equation (\ref{eq:Hamiltonian_cont}) without the $V$ term  denotes 
  the Hubbard model with a half-filled band and   
 repulsive interaction, $U$,
\cite{Solyom_adv}
 in which    the quantum fluctuation  and the spin degree of freedom 
 are discarded. 
   Equation  (\ref{eq:Hamiltonian_cont}) describes 
     the classical    charge density wave     
  where  $A =  \vf/(4\pi)$,  
  $B \sim U /a $  with Fermi velocity, $\vf$, and the lattice constant, 
 $a$, 
\cite{Suzumura_PTP79}
 The $A$ term  leads   to the uniform $\theta (x)$ while 
   the $V$ term and the $B$ term lead to spatially varying $\theta(x)$. 
 We consider the $A$ term is much larger than the $V$  term 
 and the $B$ term, i.e., 
  $A N_{imp}/L >> V$ and $A (N_{imp}/L)^2 >> B$. 
 In such a case,  the length corresponding to 
 the impurity concentration is much smaller 
  compared with the characteristic length determined by 
   $V$ and $B$ terms. 
  By taking  the average spacing of the impurity, $L/N_{imp}$, 
 as a new  lattice constant,  
  eq. (\ref{eq:Hamiltonian})  is rewritten as 
  the following  discretized model.
\cite{Suzumura_Saso}
\begin{eqnarray}
\label{eq:Hamiltonian}
{H}
 =\sum^{N_{imp}}_{j=1}\left[An_i(\theta_{j+1}-\theta_j)^2
   -V\cos(\theta_j-\zeta_j)
            -\frac{B}{n_i}\cos(2\theta_j)\right] \virg 
\end{eqnarray}
where $n_i=N_{imp}/L$ and  $\zeta_j (=-2k_F R_j)$ is a random variable
 with a Fermi momentum $\kf$. 
The characteristic length $l_0$, 
 which denotes the pinning by the impurity potential,
 is determined  by the following variational method.
 Considering  eq.(\ref{eq:Hamiltonian}) with $B = 0$, one obtains   
 the energy gain of the second term given by 
   $ - V/\sqrt{n_i l_0}$  per lattice site,  but  the increase of 
   the first term given by   $\simeq A/n_i \l_0^2$.  
Minimizing the total energy consisting of these two quantities, 
  one obtains $l_0$  as  
\cite{Fukuyama_Lee}
\begin{eqnarray}
n_il_0=\left(\frac{K_0}{\pi\epsilon}\right)^{\frac{2}{3}}
 \virg
\end{eqnarray}
 where $K_0 = \pi^2/3$ and
\begin{eqnarray}
\epsilon =  \frac{V}{4\pi An_i} 
 \point
\end{eqnarray} 
 The present paper treats the  case of small $\epsilon ( << 1 )$  
 corresponding  to  the weak impurity pinning.
 On the other hand, 
 eq. (\ref{eq:Hamiltonian}) with $V = 0$ gives the excitation 
 followed   
  by the soliton formation  with an energy, 
 $4\sqrt{2} (AB)^{1/2}$, and 
a characteristic length $d$,
\begin{eqnarray}
 d \equiv  \left( \frac{2A}{B} \right)^{1/2} \point
\end{eqnarray}
Since    $l_0$ is the pinning length  for the $V$ term and 
  $d$ is the width of the soliton for the $B$ term, 
 the competition between these two terms is characterized by 
   the quantity
\cite{Fukuyama_JPSJ78} 
\begin{eqnarray}
\overline{l_0}  =  \frac{l_0}{d(2K_0)^{\frac{1}{2}}} \point
\end{eqnarray}
It is expected that the crossover from the impurity pinning 
 to the commensurability pinning occurs for 
  $\overline{l_0} \sim o(1)$.

For the convenience of the numerical calculation, 
eq. (\ref{eq:Hamiltonian}) is rewritten   as 
\begin{eqnarray}
           \label{eq:Hamiltonian_numerical}
{H}
  =An_i\sum^{N_{imp}}_{j=1}\left[(\theta_{j+1}-\theta_j)^2
                  -4\pi\epsilon\cos(\theta_j-\zeta_j)
      - \beta \cos(2\theta_j)\right] \virg 
\end{eqnarray}
 where 
\begin{eqnarray}
 \beta = \frac{B}{A n_i^2}
   = 
 4\frac{(\pi\epsilon)^{\frac{4}{3}}}{K^{\frac{1}{3}}_0}\overline{l_0}^2 
                 \point  
\end{eqnarray}
 Equation (\ref{eq:Hamiltonian_numerical}) shows  that 
  the energy is measured in the unit of  $A n_i$ and    
    the lattice constant is taken as $1/n_i$. 
By minimizing eq. (\ref{eq:Hamiltonian_numerical}) 
 with respect to $\theta_j$, we obtain 
 the  equation for $\theta_j$  as   
\begin{eqnarray}
\label{eq:theta_j}
 (2\theta_j-\theta_{j+1}-\theta_{j-1})
      +4 \pi \epsilon \sin(\theta_j-\zeta_j)
           +2\beta\sin(2\theta_j) = 0 \virg
\end{eqnarray}
where $j = 1 , \cdots, N$.   
 A periodic boundary condition is taken where   
  $\theta_{0}=\theta_{N_{imp}}$ 
   and   $\theta_{N_{imp}+1}=\theta_1$.

The following calculation is performed by taking 
   two kinds of parameters, 
   $\epsilon$  and $\overline{l_0}$ which correspond to the variation  
       of $V$ and $B$ respectively.  

\subsection{ spatial variation of $\theta_j$}

 We calculate eq. (\ref{eq:theta_j}) 
   iteratively with several choices of initial values 
  and stop the calculation when the absolute value of the l.h.s. 
  becomes less than $10^{-7}$.
 The strength of the impurity pinning is chosen as 
    $\epsilon = 0.01$, which  
 denotes the  weak pinning and leads to the pinning  length, 
 $n_i l_0 \sim 22$.  
In this case, 
 considering the periodic boundary condition, one expects 
 that there are  
 about ten regions of $\theta_j$  exhibiting the   rapid variation
    of the order of $\pi$ 
   for a  system  with  $N_{imp} =  400$. 
 
In Fig. 1, we show the numerical results for 
 $N_{imp} =400$, $\epsilon = 0.01$, $\overline{l_0}=0, 0.5$ and  1.5
 respectively.
There are many solutions corresponding to the metastable states, 
 which  have  the energy close to the ground state.
 Figure 1  may be regarded  as the ground state since   
 the corresponding energy is  the lowest one  within 
  the present calculation. 
The variation of $\theta_j$ becomes small 
 with increasing $\overline{l_0}$ i.e., the commensurability energy. 
 Such a  state is  the solution for  the  uniform sector  
 since the spatial variation becomes  constant  (e.g. $\theta_j =0$)
 in the limit of large  $\lb0$.
\begin{figure}[htb]
\begin{center}
 \vspace{2mm}
 \leavevmode
\epsfysize=7.5cm\epsfbox{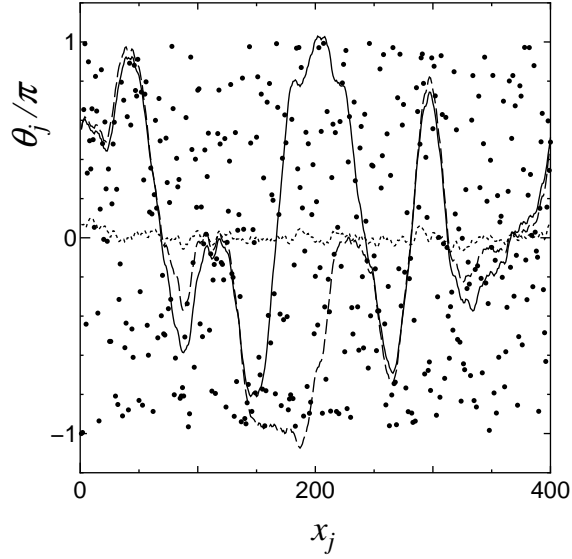}
 \vspace{-3mm}
\caption[]{
Spatial variation of $\theta_j$ for $N_{imp}=400$, $\epsilon = 0.01$
 with fixed $\overline{l_0}$ =0 (solid line), 0.5 (dashed line) 
 and 1.5 (dotted line). 
The unit of  $x_j (=1, \cdots N_{imp})$  is given by  $1/n_i$.  
 }
\end{center}
\end{figure}

Figure 2 shows  the spatial variation of $\theta_j$ 
 for the excited state in which the soliton exists. 
  For $\lb0 \gsim 1.4$,   one of the the kinks  is found at $x_j \simeq 170$ 
     although the initial value of the kink is taken at $x_j = 200$.
 This indicates that the kink is determined 
  in order to gain the energy from 
  the impurity  potential even for the large $\lb0$.
 The kink begins to move  from $x_j \simeq 170$ to $x_j \simeq 140 $
    with decreasing  $\overline{l_0}$ 
 while  the other kink also moves at lower $\lb0$ to gain the energy 
 from the impurity potential.
 However the location is fixed for the case of   $\lb0 \lsim 0.6$ 
 indicating a crossover from 
 the commensurability pinning to  the impurity pinning. 
Such a state is regarded as the solution from  the  soliton  sector 
 since the spatial variation reduces to the solution of 
 the soliton-antisoliton  in the limit of large  $\lb0$.
\begin{figure}[htb]
\begin{center}
 \vspace{2mm}
 \leavevmode
\epsfysize=7.5cm\epsfbox{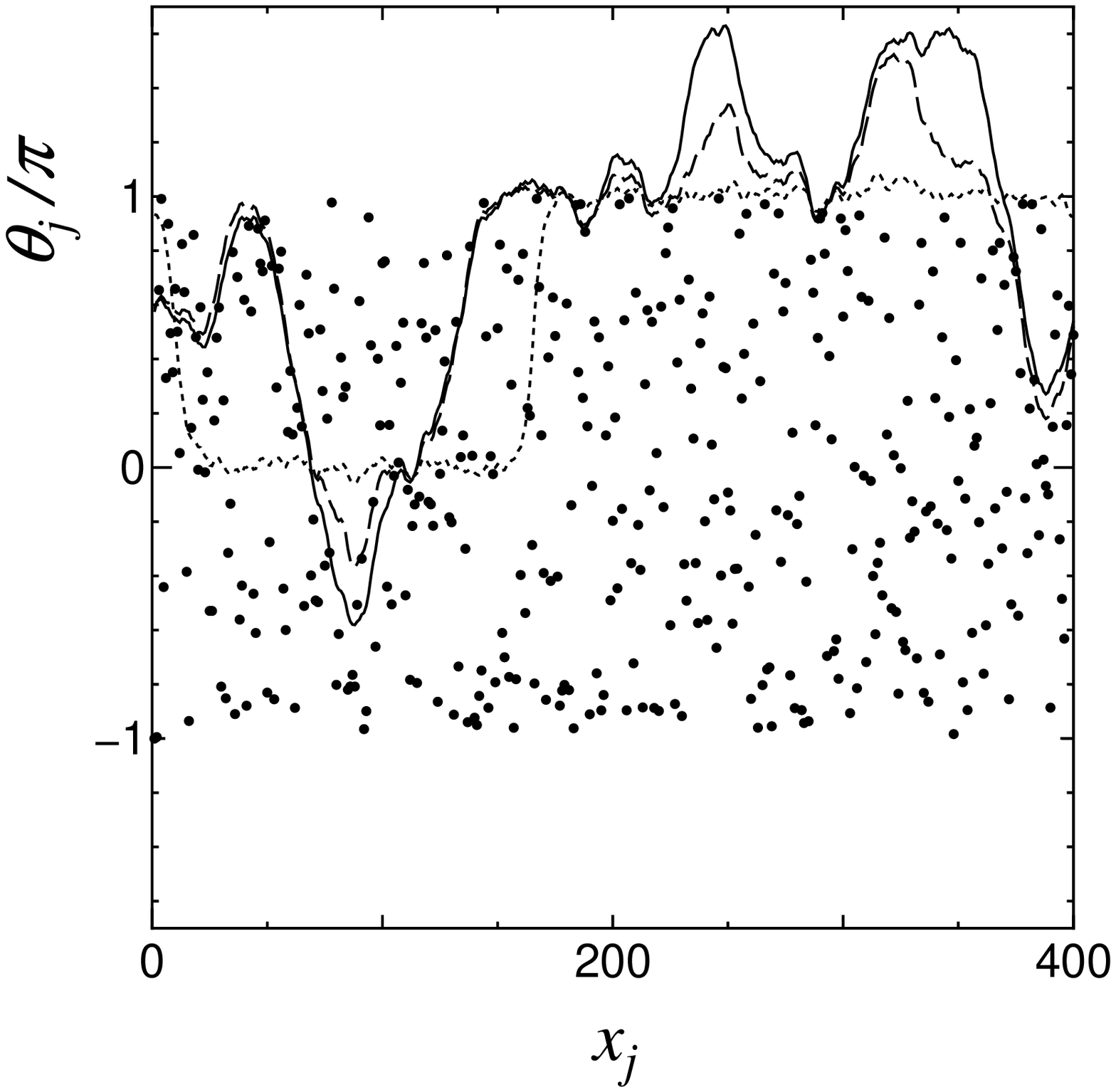}
 \vspace{-3mm}
\caption[]{
Spatial variation of $\theta_j$ for $N_{imp}=400$, $\epsilon = 0.01$
 for the excited state followed by a formation of 
 a soliton-antisoliton  where 
  $\overline{l_0}$ =0 (solid line), 0.5 (dashed line) 
 and 1.5 (dotted line). 
 }
\end{center}
\end{figure}

\subsection{excitation energy}
There are two kinds of excitations  from the ground state.
One of them is the excitation within the uniform sector, which is   
  gapless one due to the continuous distribution of the energy.
The other is the excitation followed by the soliton formation where 
 the lowest one is a state with the soliton-antisoliton.  
 The  energy difference between these two states, $\delta E_{exc}$, 
 is calculated by    
\begin{eqnarray}
\delta E_{exc}=\delta E_{sol}-\delta E_{ground} \point 
\end{eqnarray}
 Quantities 
  $\delta E_{sol}$ and $\delta E_{ground}$, which  denote 
 the deviation  of the the energy from that of the state with $\theta_j=0$, 
  are calculated by substituting  $\theta_j$  of the respective sector
 into the following equation:  
\begin{eqnarray}
\delta E& \equiv & (\frac{E}{L}+B)/ (3AK_0 / l_0^2)  
                       \nonumber  \\
  &=&\frac{1}{\pi^2N_{imp}}
  \left(\frac{\pi}{3\epsilon}\right)^{\frac{4}{3}}
       \sum^{N_{imp}}_{j=1}
 \left[ (\theta_{j+1}-\theta{j})^2
              - 4 \pi \epsilon \cos(\theta_j-\epsilon_j)
               -\beta\cos(2\theta_j) \right]
                     +\frac{4}{3}\overline{l_0}^2 
              \virg
\end{eqnarray}
where $E$ denotes an expectation value of $H$ given by 
 eq. (\ref{eq:Hamiltonian_cont}). 
 
\begin{figure}[htb]
\begin{center}
 \vspace{2mm}
 \leavevmode
\epsfysize=7.5cm\epsfbox{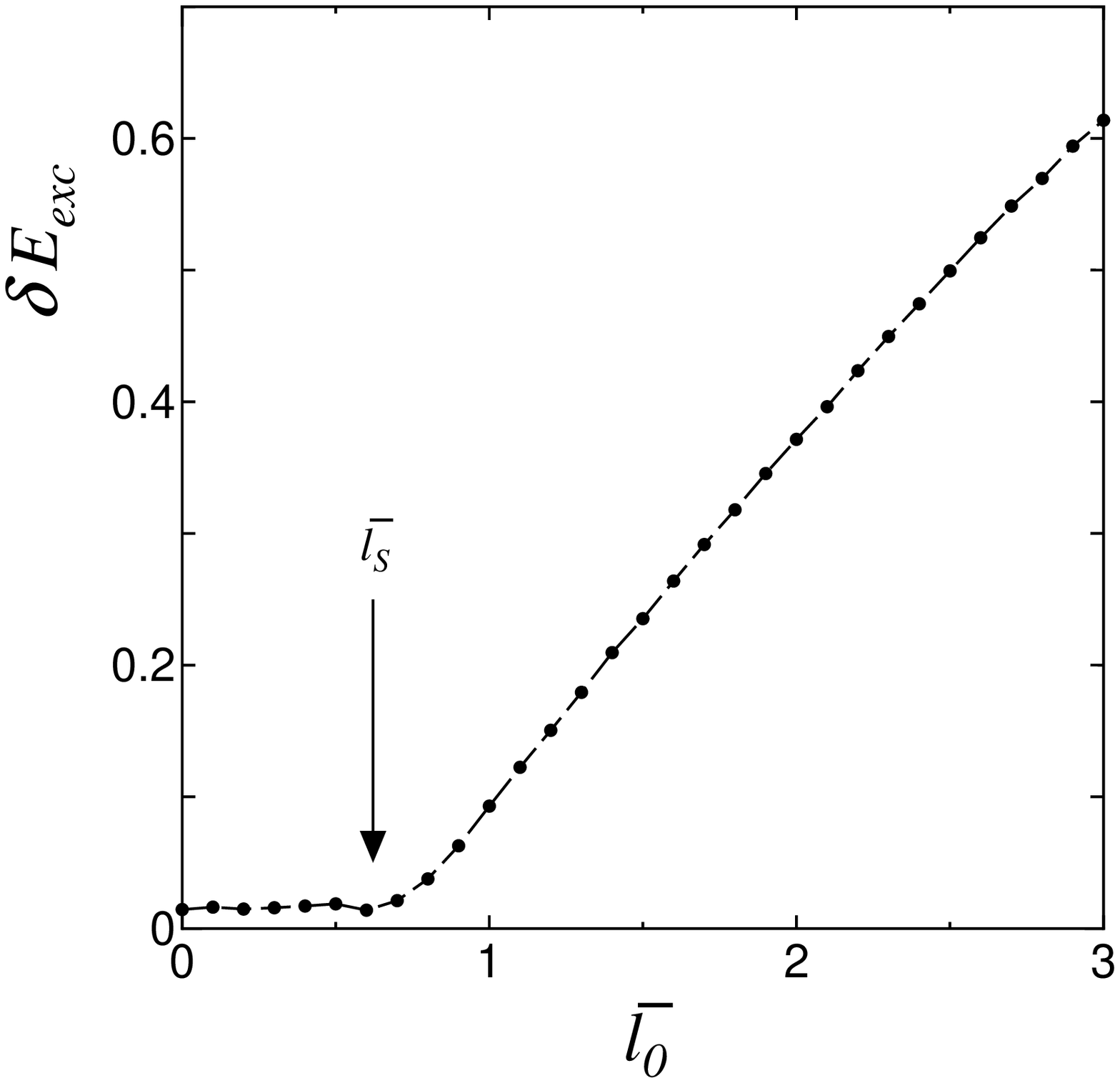}
 \vspace{-3mm}
\caption[]{
 Excited energy $\delta E_{exc}$  as a function of $\overline{l_0}$
   for $N_{imp}=400$ and  $\epsilon = 0.01$, which is averaged over 
  400 samples. 
 The arrow $\overline{l_S}$ denotes a critical value above which
 the soliton formation is followed by  a finite energy.
}
\end{center}
\end{figure}
 It should be noted that there is a critical value 
 $\overline{l_S} (\simeq  0.6) $  as shown by the arrow in Fig. 3 
where 
 $\delta E_{exc} (>0)$ is very small and almost constant 
  for $\lb0 <  \overline{l_S}$. 
Although the spatial variation of the soliton sector is apparently 
 retained  even for small $\overline{l_0}$,
  the effect of the commensurability energy is negligibly small 
   and then such a spatial variation can be regarded as one of 
         the metastable states  in the absence of the 
             commensurability energy.
For  $\lb0 \gsim \overline{l_S}$, one obtains  
 $\delta E_{exc} \sim 0.25 \lb0 - 0.8$, which indicates  
 that the formation energy of soliton-antisoliton 
  is finite but is reduced by a finite energy due to the energy gain
  by  the impurity potential.

\section{Fluctuation around the Pinned State}

In order  to find the crossover from the Mott glass to the Mott insulator,
 the degree of the local charge fluctuation 
  around the ground state 
  is examined  by calculating the response to  the external   field, 
 which gives rise to the optical conductivity.   


\subsection{local pinning frequency  $\omega_j$}
 We examine the restoring force for  $\theta_j$ 
  at the $j$-th  site when the small deviation 
   of  $\theta_j$ occurs from the equilibrium value.
 The magnitude of the force depends on the location of the lattice site 
 since the  pinning  comes from  not only the commensurate potential 
  but also the  impurity potential.
 Such a quantity is estimated  from the increase of the energy 
  by adding   the external electric field $E'$ which 
  gives an additional term, $- E \theta$,  to  the r.h.s.  of 
   eq.(\ref{eq:Hamiltonian_cont}) where $E = e E'/\pi$ with 
     $e$ being the electric charge.
 \cite{Fukuyama_JPSJ76} 
 In this case, the lattice model is rewritten as   
\begin{eqnarray}
 \label{eq:Hamiltonian_lattice_E}
H  = An_i\sum^{N_{imp}}_{j=1}\left[(\theta_{j+1}-\theta_j)^2
   - 4\pi\epsilon\cos(\theta_j-\zeta_j) 
 -  \beta \cos(2\theta_j)
       - \gamma  \overline{E}\theta_j\right] 
            \virg
\end{eqnarray}
  where 
 $\overline{E}= \overline{l_0}^2/(3AK_0)$ and 
 \begin{eqnarray}
\gamma\equiv\frac{3(\pi\epsilon)^\frac{4}{3}}{K_0^{\frac{1}{3}}}
 \point
\end{eqnarray}
The resultant equation for  $\theta_j$ is given by  
\begin{eqnarray} 
 \label{eq:sce_E}
 (2 \theta_j-\theta_{j+1}-\theta_{j-1})
    +\alpha\sin(\theta_j-\zeta_j)
   +2\beta\sin(2\theta_j)-\gamma\overline{E} = 0 \point
\end{eqnarray}
By using $\theta_j(E)$ of the solution of eq. (\ref{eq:sce_E}),
   we calculate 
\begin{eqnarray}
 1 / \omega_j^2  =   
\lim_{\overline{E} \rightarrow 0}
 \left( \theta_j(E) - \theta_j(0)  \right) / \gamma \overline{E} 
          \point 
\label{eq:omega_j}
\end{eqnarray}
The quantity $\omega_j$ is the local pinning frequency 
 since the increase of the energy in the presence of the
 small external field can be expressed as  
 $ \omega_j^2  (\theta_j(E) - \theta_j(0))^2 /2$ ,
  which is calculated  by expanding 
 eq. (\ref{eq:Hamiltonian_lattice_E})
  in terms of $\theta_j(E)-\theta(0)$.

In Fig. 4, $\lb0$ dependence of $<\omega_j>$  
 is shown for  $N_{imp}=4000$ and  $\epsilon = 0.01$ 
 where $<>$ denotes the spatial average 
 and the closed square is the result   
   averaged over 20 samples.
 The calculation of  $\omega_j$ has been performed 
  by choosing  $\overline{E}=0.001$, which 
 is small enough  to calculate  eq. (\ref{eq:omega_j}).
 For $\lb0  \lsim 0.4$, 
 the quantity $<\omega_j>$ is constant indicating 
 no visible effect  of  the  commensurability pinning. 
 The commensurability pinning dominates for  $\lb0 \gsim  0.6$,
 in which    $\omega_j \propto \lb0$.
 It seems that a crossover from the impurity pinning to 
 the commensurability pinning occurs in the interval region 
  of  $ 0.4 \lsim \lb0 \lsim  0.6$. 
\begin{figure}[htb]
\begin{center}
 \vspace{2mm}
 \leavevmode
\epsfysize=7.5cm\epsfbox{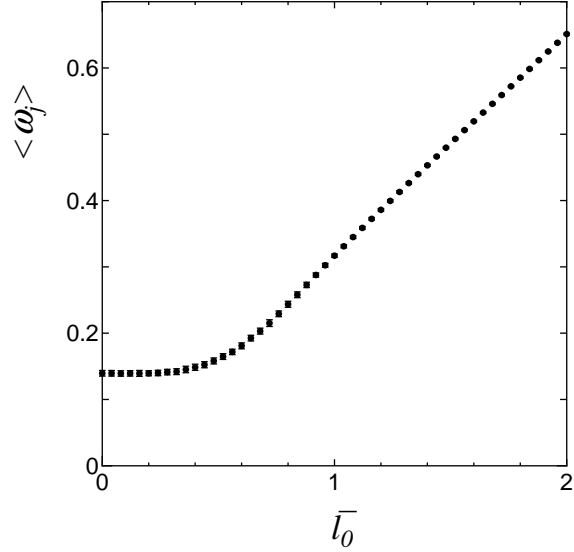}
 \vspace{-3mm}
\caption[]{
$\overline{l_0}$ dependence of $< \omega_j >$ 
 for $N_{imp}=4000$ and  $\epsilon = 0.01$, 
  which is averaged over 20 samples. The limiting behavior 
 is evaluated by taking $\overline{E}=0.001$.
 The error bar becomes negligibly small in the visible scale 
  with increasing $\overline{l_0}$.
}
\end{center}
\end{figure}


\subsection{root mean square  of $\omega_j$}
The property of the spatial variation of $\omega_j$ 
 is examined by calculating   the  root mean square  
 of $\omega_j $.
 The normalized root mean square  of $\omega_j$ is given by
\begin{eqnarray}
  z = \frac{1}{\langle\omega_j\rangle}
     \sqrt{\langle(\omega_j-\langle\omega_j\rangle)^2\rangle}
           \point 
\end{eqnarray}
 The larger $z$  indicates the broader  distribution of 
  the pinning force.    
It is expected that z  of the impurity pinning 
 is much larger than that of the commensurability pinning. 
The fluctuation of  $z$ is large 
 for the choice of the present sample with $N_{imp}$ 
 where 
 $z_s$ is used to distinguish the respective sample.
Thus, we  further  examine the root mean square of $z_s$  
 given by 
\begin{eqnarray}
\sqrt{\langle(z_s-\langle z_s\rangle)^2\rangle}
          \point 
 \end{eqnarray}
 
\begin{figure}[htb]
\begin{center}
 \vspace{2mm}
 \leavevmode
\epsfysize=7.5cm\epsfbox{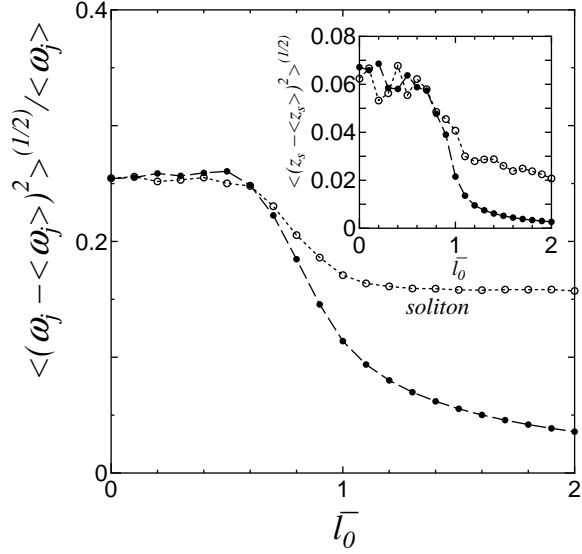}
 \vspace{-3mm}
\caption[]{
$\overline{l_0}$ dependence of the normalized root mean square 
 of $\omega_j$ for 
  $N_{imp}=400$,  $\epsilon = 0.01$ and  $\overline{E}=0.001$ 
  which is averaged over 400 samples. 
 The closed circle and solid line (open circle  and dotted line)
 denote the results for the ground state (the excited state with soliton).
The inset denotes  the corresponding 
  root mean square of $z_s$  given by 
$\sqrt{\langle(z_s-\langle z_s\rangle)^2\rangle}$.
}
\end{center}
\end{figure}
Figure 5 (dashed line and closed circle )
  shows the normalized root mean square 
   of $<\omega_j>$, i.e. $<z_s>_s$  
 for  $N_{imp}=400$,  $\epsilon = 0.01$ and  $\overline{E}=0.001$ 
  where  400 samples are used for $s$.
For $\lb0 \lsim   0.6$,  $<z_s>_s$ is almost constant 
 while it decreases for   $\lb0 \gsim   0.6$ due to 
 the increase of the commensurability pinning. 
The result for the soliton sector  is  shown by the open circle 
 where it   remains finite even for large $\lb0$.
 This is understood as follows. 
 When the impurity is absent, 
  the location of the kink is easy to move under the external field 
  resulting in  the extremely small $\omega_j$.  
  However the presence of the impurity pinning  
   prevents  such a movement  leading to finite $\omega_j$, which  
    gives  the variation similar to the impurity pinning.    
The deviation from that of the uniform sector is found for 
 $\lb0 \gsim  0.7$, which is  slightly larger 
  than $\overline{l_S}$.  
 This indicates a possibility that the fluctuation around 
   the pinned state   even for $\lb0 \gsim \overline{l_S}$
    still resembles  that of $\lb0 = 0$.  
 The inset shows the root mean square  of such $z_s$  which 
  gives a critical value  $\lb0 \simeq  1.0$, above which  
    the  spatial fluctuation  of $\omega_j$ decreases rapidly
    indicating a crossover  into the Mott insulating state.

This tendency  can be also seen from 
 the distribution of $\omega_j$, which is defined by 
\begin{eqnarray}
P(\omega)=\frac{1}{N_{imp}}\sum^{N_{imp}}_{j=1}\delta(\omega-\omega_j)
          \point 
\end{eqnarray}
In  Fig. 6, 
$\omega (= \omega_j)$ dependence of  $P$ is shown for 
    $N_{imp}=4000$,  $\epsilon = 0.01$ and  $\overline{E}=0.001$
 where $\overline{l_0}$ = 0 (closed circle), 0.5 (open circle), 
 0.8 (closed triangle) and 1.5 (closed diamond).  
  The average over   20 samples are used 
  for the numerical accuracy of small $\omega$. 
The quantity $P$ with $\lb0=0$ is almost the same
 as $P$ with $\lb0 = 0.5$ and  still resembles  $P$ with $\lb0 = 0.8$
 while it is much different from $P$ with $\lb0 = 1.5$.
 This indicates that   the distribution changes the property  
  for   $\lb0 >  \overline{l_G}$ , 
  where $\overline{l_G}$ is larger than at least 0.8. 
 

\begin{figure}[htb]
\begin{center}
 \vspace{2mm}
 \leavevmode
\epsfysize=7.5cm\epsfbox{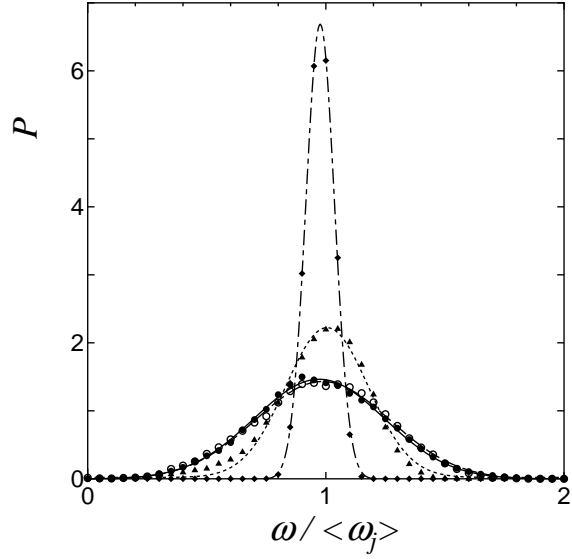}
 \vspace{-3mm}
\caption[]{
Normalized distribution of $\omega_j$, $P(\omega/<\omega_j>)$,
 for  
   $N_{imp}=4000$,  $\epsilon = 0.01$ and  $\overline{E}=0.001$
 where $\overline{l_0}$ = 0 (closed circle), 0.5 (open circle), 
 0.8 (closed triangle) and 1.5 (closed diamond).  
  The results are obtained by  averaging  over 20 samples. 
}
\end{center}
\end{figure}

\subsection{optical conductivity}
Based on Figs. 3, 4, 5 and 6, 
 it is reasonable to admit  that  the intermediate region 
 with $\overline{l_S} < \lb0 < \overline{l_G}$  
   may correspond to that of the Mott glass.
 However the estimation for $\overline{l_G}$ is rather complicated 
 compared with  $\overline{l_S}$. 
 Then  we  examine  the optical conductivity
  since the Mott glass is maintained 
 to exhibit the  optical  conductivity given  by 
   $\sigma (\omega) \propto \omega^2$.
\cite{Orignac_PRL} 
We calculate the conductivity within the classical treatment,
 by noting the fact that 
 the electric dipole moment  at the $j$-th site  is given by 
 $ (2 \kf / \pi) e y_j  = (e / \pi) \theta_j   
$
 with $y_j$ being the displacement.
\cite{Fukuyama_JPSJ76}
 In the presence  of the electric 
 field $E_0 \e^{i \omega t}$, the equation of 
 motion for the charge  density wave  is given by 
 \begin{eqnarray}
 m^*\frac{d^2y_j}{dt^2}
  =-\frac{m^*}{\tau}\frac{d y_j}{dt}+eE_0 \e^{i \omega t} 
           - m^*\omega_j^2 y_j \virg 
\end{eqnarray}
where $m^*$ is the effective mass and  the life time, $\tau$, 
 is introduced 
 for the convergence of the numerical calculation.
 The quantity $\omega_j$, which corresponds to  the  restoring force,   
 $ m^* \omega_j^2 y_j$ is calculated from 
  eq. (\ref{eq:omega_j}). 
The quantity $y_j$ is calculated as  
\begin{eqnarray}
y_j = \frac{eE_0 / (i \omega m^*)}
{\frac{1}{\tau}
  +i  \omega\left(1-\frac{\omega_j^2}{\omega^2}\right)}
 \virg 
\end{eqnarray}
 which leads to the local conductivity at the $j$-th site 
\begin{eqnarray}
\sigma_j(\omega) 
   =\frac{ 2 \kf e^2\tau}{m^*\pi}\frac{1}
         {1+i\omega\tau\left(1-\frac{\omega_j^2}{\omega^2}\right)}
 \point
\end{eqnarray}
 Thus one obtains the real part of the total conductivity as 
\begin{eqnarray}
  \label{eq:total_con}
\sigma(\omega) = 
 \frac{2\tau}{\pi N_{imp}}\sum_{j=1}^{N_{imp}}\frac{1}{1+(\omega\tau)^2\left(1-\frac{\omega_j^2}{\omega^2}\right)^2} \virg
\end{eqnarray}
where the conductivity is normalized such that  
 the summation of $\sigma(\omega)$ with respect to 
 $\omega$ becomes equal to unity.
In Fig. 7,  optical conductivity is shown 
for 
  $N_{imp}=4000$,  $\epsilon = 0.01$ and  $\tau = 20$
 where respective results are for
$\overline{l_0}$ = 0 (closed circle), 0.5 (open circle), 
 0.8 (closed ) and 1.5 (closed diamond). 
\begin{figure}[htb]
\begin{center}
 \vspace{2mm}
 \leavevmode
\epsfysize=7.5cm\epsfbox{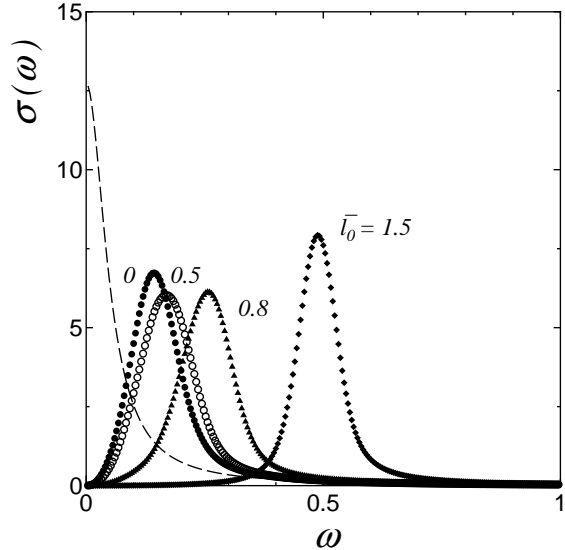}
 \vspace{-3mm}
\caption[]{
Normalized optical conductivity for  
   $N_{imp}=400$,  $\epsilon = 0.01$ and  $\tau = 20$,
     which are averaged over 100 samples.
 Respective results are for
$\overline{l_0}$ = 0 (closed circle), 0.5 (open circle), 
 0.8 (closed ) and 1.5 (closed diamond).  
 The dashed curve corresponds to the conductivity of 
   the Drude's formula (i.e., $\omega_j \rightarrow 0$). 
}
\end{center}
\end{figure}
The conductivity for $\lb0 = 0$ denotes  the  
 conventional impurity pinning of the charge density wave
 where the frequency  of $\omega$ corresponding to  the peak 
  is essentially the same as $< \omega_j >$, i.e., 
   the average of $\omega_j$ shown in Fig. 4.
 Such  a frequency  increases with increasing 
  $\lb0$, which is consistent with the result of Fig. 4. 
 The width of $\omega$ is determined by the root mean square 
  of $\omega_j$,  (Fig. 5).   
 The width for small $\lb0$, 
  which is determined by the distribution of the pinning force,  
   decreases for large  $\lb0$ 
  due to the increase of the commensurability pinning. 
 It is found that  the peak of $\omega$ is slightly reduced  
   in region of the Mott glass
 ($\overline{l_S} < \lb0 < \overline{l_G}$), 
 since  the width of $\sigma (\omega)$ in the Mott glass   becomes large 
 due to the competition between the impurity pinning 
 and the commensurability pinning. 
For comparison, the conductivity of the Drude's formula
 is shown by the dashed curve where the width is 
chosen as $\tau =20$. 
 The conductivity for  $\lb0 =$0, 0.5 and 0.8 
 exhibits $\sigma (\omega) \propto \omega^2$ for small $\omega$ 
  within the 
  numerical accuracy of the present calculation 
 indicating  the property of the Anderson glass.
In order to examine   such a region, 
 we calculate  a coefficient, $A$,  
of  $\sigma (\omega)$ which is given  
  in the limit of small $\omega$ by        
\begin{eqnarray}
 \label{eq:A}
 \sigma (\omega) \simeq A \omega^2
\point
\end{eqnarray}
   In Fig. 8, the quantity $A$ 
 is shown by taking   
  $\omega \rightarrow 0.01$,  
 where the dashed line denotes the guide for the eye.
    The inset depicts the procedure  to calculate 
 $A$ by an extrapolation of $A(\tau)$ for $1/\tau =$ 1, 0.1 and 0.05
 since eq. (\ref{eq:A}) depends on $\tau$.  
 With increasing $\lb0$, 
 the quantity $A$ stays almost constatnt for small $\lb0$
  while it  decreases rapidly  
    for $\lb0 \simeq 1.0$.  
    However, the data  for $\lb0 >1.0$ is beyond  
     the present  calculation, which is based on  
          the model of the classical charge density wave,  and  
   eq. (\ref{eq:total_con}) showing  the gradual 
       decrease of $A$  even in the Mott insulating phase.  
  Thus  we concludes  $\overline{l_G} \simeq  1.0$, 
    which is shown by the arrow in the main figure.    
\begin{figure}[htb]
\begin{center}
 \vspace{2mm}
 \leavevmode
\epsfysize=7.5cm\epsfbox{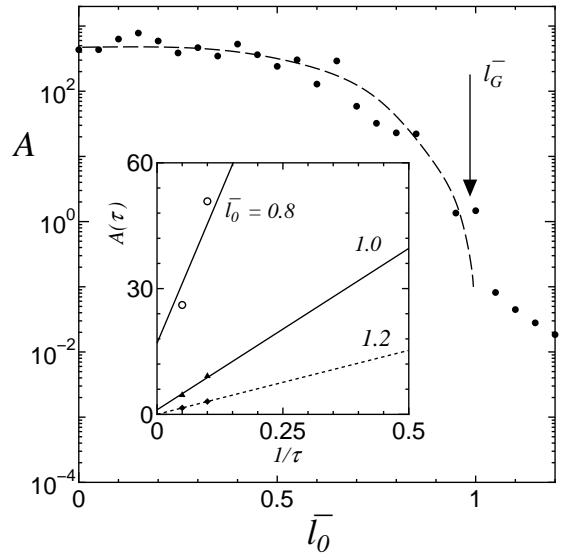}
 \vspace{-3mm}
\caption[]{
$\overline{l_0}$ dependence of $A$, 
 which denotes $\sigma (\omega) \simeq A \omega^2$ 
 in the limit of small $\omega$. 
In the inset, $A$ is obtained from the value of the vertical axis
   where  the  extrapolation for $1/\tau$ is shown 
   for $\lb0 =$ 0.8. 1.0 and 1.2.
 The steep decrease of $A$ leads to 
 $\overline{l_G}  \simeq 1.0$ as shown by the arrow.  
}
\end{center}
\end{figure}


\section{Discussion}
We have examined the competition between the impurity pining 
 and the commensurability pinning 
  of the charge density wave, 
 which leads to  three kinds of phases as a crossover. 
 It is found  that  the Mott glass state exists  
 in the intermediate  region of  $\overline{l_S} < \lb0 < \overline{l_G}$
 with   $\overline{l_S} \simeq 0.6$ and 
   $ \overline{l_G} \simeq 1.0$ 
 while there are the region of 
  the Anderson glass for $\lb0  < \overline{l_S}$
   and that of the Mott insulator for $\overline{l_G} < \lb0$. 
 
The intermediate region corresponding to the Mott glass 
seems to be  consistent with that  of OGD,  
\cite{Orignac_PRL}
 who obtained  $0.98 <  d/\overline{l_0} < 1.58$ in their 
  notations.  
 The Mott glass undergoes 
  the crossover to the Anderson glass (or Mott insulator) 
  with decreasing ( or increasing) $\lb0$
  due to the effect of the impurity, which has the wide distribution 
  of the pinning force (i.e. restoring force).
   Then there is    a tiny domain  of the soliton formation 
 in the state of  the Mott glass. 
 Even in the presence of  
  such a state,   the  optical conductivity 
   similar to the Anderson glass could be expected
 due to the contribution from the slowly varying region
 \cite{Orignac_cond}
  since the conductivity  
    originates in the distribution of the restoring force
     in  the uniform sector. 
 The  conductivity  with low frequency resembles 
 that of the Anderson glass, which denotes    
    no  gap in the conductivity, i.e.,  
  $\sigma (\omega) \propto \omega^2$  for small $\omega$. 
 Note that it is  beyond the present calculation to 
   examine if the conductivity exhibits a property 
 $\sigma (\omega) \propto \omega^2 {\rm ln}^2(\omega \tau)$.
\cite{Impurity1,Impurity2}
   The  existence  of the  gap  for the formation of the phase soliton,
 which leads to the variation of the electron number,     
  is equivalent to 
   the  absence of compressibility since   
     a finite energy is needed to  add or remove  an electron.  
 However  the case of doping of electron or hole, 
 which corresponds to the ground state
  with the soliton formation,     leads to  
  the absence of the Mott insulator and   
     the optical conductivity  with no gap  
        as expected from the soliton sector of Fig. 5.  
   
Here we comment on the present mode of  the  charge density wave, 
 which is relevant to  the electron system. 
  The latter case includes  two additional factors,
 i.e, the  quantum fluctuation and the spin fluctuation.
 These factors have at least the  effect of reducing the magnitude of 
 $V$ and $B$ in eq. (\ref{eq:Hamiltonian}), which could be 
  taken  into account by adjusting 
  parameters. However the novel  effect can be  expect 
  when the depinning begins. 
   This problem will be discussed in a separate paper. 



\begin{thebibliography}{99} 
\def\jo #1#2#3#4{#1 {\bf #2} (#3) #4}  
\def\JPSJ{J.\ Phys.\ Soc.\ Jpn.}
\def\PRB{Phys.\ Rev.\ B}
\def\PRL{Phys.\ Rev.\ Lett}
\def\PTP{Prog.\ Theor.\ Phys.}
\def\ADV{Adv.\ Phys.}

\bibitem{Impurity1}
V.L. Berezinskii:
Zh. Eksp. Teor. Fiz. {\bf 65}(1973)1251
[Sov. Phys.-JETP {\bf38} (1974) 620].

\bibitem{Impurity2}
M.V.  Feigelman and Vinokur: Solid State Commun.
 {\bf 45} (1983) 603

\bibitem{localization}
E. Abraham, P.W. Anderson, D.C. Licciardello and
 T.V. Ramakrishnan: 
 \jo{\PRL}{42}{1979}{673}.


\bibitem{Suzumura_Fukuyama_JPSJ84}
 Y. Suzumura and H. Fukuyama:
\jo{\JPSJ}{53}{1984}{3918}.

\bibitem{Giamarchi_PRB1988}
 T. Giamarchi and H.J. Schulz:
\jo{\PRB}{37}{1988}{325}.

\bibitem{Fukuyama_Lee}
 H. Fukuyama and P.A. Lee:
\jo{\PRB}{17}{1978}{535}.

\bibitem{Fukuyama_JPSJ78}
 H. Fukuyama:
\jo{\JPSJ}{45}{1978}{1474}.

\bibitem{Fujimoto_Kawakami_PRB1996}
 S. Fujimoto  and N. Kawakami:
\jo{\PRB}{54}{1996}{R11018}.

\bibitem{Mori_Fukuyama_JPSJ1996}
 M. Mori and H. Fukuyama:
\jo{\JPSJ}{65}{1996}{3604}.

\bibitem{Orignac_PRL}
 E. Orignac, T. Giamarchi and P. Le Doussal:
 \jo{\PRL}{83}{1999}{2378}.
\bibitem{Fujimoto_JPSJ2000}
 S. Fujimoto:
\jo{\JPSJ}{69}{2000}{2597}.

\bibitem{Suzumura_Meeting}
Y. Suzumura and M. Isobe:
 read at  Meeting of Physical Society of Japan,
 Sendai, March 28-31, 2003.
 

\bibitem{Suzumura_PTP79}
 Y. Suzumura:
\jo{\PTP}{61}{1979}{1}.

\bibitem{Solyom_adv}
 J. S\'olyom:
\jo{\ADV}{28}{1979}{201}.

\bibitem{Suzumura_Saso}
 Y. Suzumura and T. Saso:
\jo{\JPSJ}{55}{1986}{4359}.

\bibitem{Fukuyama_JPSJ76}
 H. Fukuyama:
\jo{\JPSJ}{41}{1976}{513}.

\bibitem{Orignac_cond}
T. Giamarchi, P. Le Doussal and  E. Orignac: 
\jo{\PRB}{64}{2001}{245119}.





\end{thebibliography}
\end{document}